\documentclass[usenatbib]{mnras}

\usepackage[T1]{fontenc}
\usepackage{ae,aecompl}

\usepackage{graphicx}	
\usepackage{amsmath}	
\usepackage{amssymb}	

\newcommand{\nc}{\newcommand}
\nc{\teff}{$T_{\rm eff}$\,}
\nc{\logg}{\rm log\,$g$\,}
\nc{\kms}{\,${\rm km\,s}^{-1}$\,}
\nc{\mic}{$\xi_{\rm t}$\,}
\nc{\ms}{~m~s$^{-1}$\,}

\title[First Super-Earth from Dharma Planet Survey]{The first super-Earth Detection from the High Cadence and High Radial Velocity Precision Dharma Planet Survey}

\author[Ma et al.]{
Bo Ma$^{1}$\thanks{E-mail: boma@ufl.edu}, Jian Ge$^{1}$, Matthew Muterspaugh$^{2,3}$, Michael A. Singer$^{1}$, Gregory W. Henry$^{3}$, 
\newauthor Jonay~I. Gonz\'alez Hern\'andez$^{4,5}$, Sirinrat Sithajan$^{1}$, Sarik Jeram$^{1}$,  Michael Williamson$^{3}$, 
\newauthor Keivan Stassun$^{6}$,  Benjamin Kimock$^{1}$, Frank Varosi$^{1}$, Sidney Schofield$^{1}$, Jian Liu$^{1}$, 
\newauthor Scott Powell$^{1}$, Anthony Cassette$^{1}$, Hali Jakeman$^{1}$, Louis Avner$^{1}$, Nolan Grieves$^{1}$, 
\newauthor Rory Barnes$^{7}$, Sankalp Gilda$^{1}$, Jim Grantham$^{8}$, Greg Stafford$^{8}$, David Savage$^{8}$, 
\newauthor Steve Bland$^{8}$, and Brent Ealey$^{8}$ \\
$^{1}$Department of Astronomy, University of Florida, 211 Bryant Space Science Center, Gainesville, FL, 32611-2055, USA \\
$^{2}$Department of Mathematics and Physics, College of Life and Physical Sciences, Tennessee State University, Boswell Science Hall, \\ 
Nashville, TN 37209, USA \\
$^{3}$Tennessee State University, Center of Excellence in Information Systems, Nashville, TN 37203, USA \\
$^{4}$Instituto de Astrof\'{\i}sica de Canarias, C/V\'{\i}a L\'actea S/N, E-38200 La Laguna, Spain \\
$^{5}$Departamento de Astrof\'{\i}sica, Universidad de La Laguna, E-38205 La Laguna, Tenerife, Spain \\
$^{6}$Department of Physics \& Astronomy, Vanderbilt University, Nashville, TN 37235 USA \\
$^{7}$Astronomy Department, University of Washington, Box 951580, Seattle, WA 98195, USA \\
$^{8}$Steward Observatory, The University of Arizona, Tucson, AZ, 85719, USA
}

\date{Accepted XXX. Received YYY; in original form ZZZ}

\pubyear{2018}

\begin{document}
\label{firstpage}
\pagerange{\pageref{firstpage}--\pageref{lastpage}}
\maketitle

\begin{abstract}
The Dharma Planet Survey (DPS) aims to monitor about 150 nearby very bright FGKM dwarfs (within 50 pc) 
during 2016$-$2020 for low-mass planet detection and characterization using the TOU very high resolution
optical spectrograph (R$\approx$100,000, 380-900nm). TOU was initially mounted to the 2-m Automatic Spectroscopic 
Telescope at Fairborn Observatory in 2013-2015 to conduct a pilot survey, then moved to the dedicated 50-inch automatic 
telescope on Mt. Lemmon in 2016 to launch the survey. Here we report the first planet detection from DPS, 
a super-Earth candidate orbiting a bright K dwarf star, HD~26965. 
It is the second brightest star ($V=4.4$~mag) on the sky with a super-Earth candidate. The planet candidate 
has a mass of 8.47$\pm0.47M_{\rm Earth}$, period of $42.38\pm0.01$~d, and eccentricity of $0.04^{+0.05}_{-0.03}$. 
This RV signal was independently detected by Diaz et al. (2018), but they could not confirm if the signal is from a 
planet or from stellar activity. The orbital period of the planet is close to the rotation period of the star (39$-$44.5~d) 
measured from stellar activity indicators. Our high precision photometric campaign and line bisector analysis of 
this star do not find any significant variations at the orbital period. Stellar RV jitters modeled from star spots and convection 
inhibition are also not strong enough to explain the RV signal detected. After further comparing RV data from the star's 
active magnetic phase and quiet magnetic phase, we conclude that the RV signal is due to planetary-reflex motion and not stellar activity.  

\end{abstract}




\section{Introduction}

Results emerging from the Kepler mission and ground-based radial velocity (RV) surveys 
reveal a population of close-in low-mass planets orbiting FGKM stars \citep{howard10, howard12, mayor11, bonfils13}. 
Most of these low-mass planets have orbital periods shorter than the 88-day orbit of Mercury, and
 many of them are in very compact multiple-planet systems \citep[e.g.][]{howard12, bat13, mul15, cou16, mor16}. 
 This unexpected population of close-in low-mass planets (super-Earths and Neptune-mass planets), which is completely 
absent in our own Solar System, is surprisingly common and represents the most 
dominant class of planetary systems known to date. However, the measured occurrence
 rate of this close-in low-mass planet population varies significantly between different RV groups, ranging from 
 $\sim$23\% to 50\% \citep{howard10, mayor11, bonfils13}. 
The large uncertainties in the ground-based RV survey results are largely
 due to their low-cadence survey strategy.
 For instance, the average number of RV measurements per survey target is between 20-40, 
 and they were typically spread out over six years \citep{howard10, mayor11, bonfils13, motalebi15, bor17, per17}.

 Kepler has enabled estimates of the occurrence rates
 of low-mass planets with orbits as long as 300 days \citep[e.g.][]{pet13, for14, burk15}. However, the uncertainties in the Kepler 
measurements remain large due to the unknown false positive rate, and systematic errors caused by
 Kepler's pipeline completeness, survey selection effects, and catalog reliability \citep{burk15, chris16}. For instance, the estimated false 
positive rate is $\sim$11\% for low mass planet candidates \citep{fre13} 
and $\sim$55\% for giant planet candidates \citep{san16}.
 The systematic errors in candidate detection have led to a factor of 2-3 times 
difference in estimates of the occurrence rates of low-mass exoplanets
 from different transiting groups \citep[e.g.][]{howard12, pet13, for14, dressing15, burk15, mulders15}. In addition, due to strict edge-on 
 geometry requirements, Kepler may have missed some non-transiting planets in the Kepler transit planet 
systems, leading to additional uncertainties in the occurrence rate 
measurements \citep[e.g.,][]{buchhave16}.

It is quite clear that an independent and uniform measurement of the
 occurrence rate of this close-in small planet population is necessary. 
This independent survey  will not only help address the discrepancies between previous surveys and constrain 
planet formation theories, but also help independently resolve controversial low-mass 
planet discovery claims by different RV surveys using different RV instruments, or 
different data pipelines using the same instrument. For example, two groups reported four \citep{mayor09} 
and six \citep{vogt10} low mass planets orbiting  GJ 581, respectively. The same two groups reported 
six planets \citep{vogt15} and four planets \citep{motalebi15} orbiting HD 219134, respectively. 
These uncertainties greatly affect our understanding of
 exoplanet systems and their architectures, especially those with low-mass planets.
 More independent observations from high-precision RV campaigns are required to resolve these debates.

 High cadence and high RV precision observation of survey stars can significantly
 improve sensitivity for detecting close-in low-mass planets.
 It is very challenging to search for close-in low-mass planets which produce 
very small RV signals over 1-2 month periods, as previous RV surveys on large telescopes 
(such as Keck and HARPS) often suffer from sparse and irregular observation cadences due to 
sharing requirements. For example, \citet{ang16} pointed out that uneven and sparse sampling is one of the 
reasons why Proxima b could not be unambiguously confirmed with their pre-2016 RV data.
This likely accounted for the large discrepancy in low-mass planet 
occurrence rates by different groups \citep{howard10, mayor11}. 
Continuous phase coverage with high RV precision would likely remove these discrepancies. 
Pioneering observations by HARPS of 10 very stable FGK dwarfs with high cadence 
($\sim$50 data points per observing season, and an average of 122 RV points per star)
 and precision ($\sim$0.9-2.6 m/s) led to detection of three low-mass planetary systems
 with 6 low-mass planets \citep[one with as low as 3.6 M$_{\earth}$,][]{pepe11}, 
which otherwise would have largely escaped detection.

The Dharma Planet Survey (DPS) was designed to detect and 
characterize close-in low-mass planets and sub-Jovian planets.  
The ultimate survey goal is to detect potentially habitable super-Earth planet 
candidates and provide bright high-priority follow-up targets for future
 space missions (such as JWST, WFIRST-AFTA, EXO-C, EXO-S, and LUVOIR surveyor) 
to identify possible biomarkers supporting life \citep{ge16}. It will initially
 search for and characterize low-mass planets around $150$ nearby very bright FGKM dwarfs 
in 2016-2020.

%

  The DPS survey, unlike previous and on-going low-mass high-precision RV planet
surveys with varying numbers of measurements \citep[from a few RV data points to $\sim$400 RV data points, e.g.,][]{dumusque12}, 
will offer a nearly homogeneous high cadence for every survey target. Every target 
will be initially observed $\sim$30 consecutive observable nights to target close-in
 low-mass planets detection. After that, each target will be observed an additional 
$\sim$70 times randomly spread over 420 days.
The automatic nature of the 50-inch telescope and its flexible queue observation 
schedule are key to realizing this nearly homogenous high cadence. 
This cadence will minimize time-aliasing and RV jitter effects caused by stellar
 activity that often preclude the detection of low-mass planets, 
especially those in highly eccentric orbits, which may have been missed
 by previous surveys \citep{dumusque11a, vanderburg16}. Because of the high cadence for every survey star, 
both detections and non-detections from the survey can be reliably used
 for statistical studies. The proposed survey strategy, cadence, 
and schedule will therefore offer the optimal accuracy to assess the survey 
completeness and to determine occurrence rates of low-mass planets. 
This survey will offer a homogeneous data set for constraining formation models
 of low-mass planets with periods less than 450 days.
  In addition, this DPS survey strategy provides an efficient way to explore habitable
 low-mass planets around nearby FGKM dwarfs with greatly improved survey sensitivity
 and completeness compared to previous Doppler surveys. 

In this paper, we report the first planet detection from DPS survey, a super-Earth candidate orbiting 
a nearby bright K0.5V star with $V=4.4$~mag, HD~26965. The RV signal has also been reported recently 
in \citet{diaz18}, in which they claim it is either a planet signal, or a signal from stellar activity. 
In section~2, we describe the observations used in this paper. We present stellar parameters for the star 
in section~3 and orbital parameters for the planet candidate in section~4. We discuss the nature of the 
radial velocity signal in section~5. In section~6 we discuss our results and present our conclusions. 

\section {Observations and Radial Velocity Extraction}
\subsection{TOU RV Data}
TOU (formerly called EXPERT-III) is a fiber-fed, cross-dispersed echelle 
spectrograph with a spectral resolution of about 100,000, 
wavelength coverage of $3800-9000$\AA, and a 4kx4k Fairchild CCD detector \citep{ge12,ge14}. 
The instrument holds a very high vacuum of 1 micro torr and about 1mK 
temperature stability over a month. 

We obtained 66 observations of HD~26965 using TOU  at 
the 2-m Automatic Spectroscopic Telescope (AST) at Fairborn Observatory between 
2014 and 2015. We later moved TOU to the UF 50-inch robotic telescope at Mt. Lemmon, 
called the Dharma Endowment Foundation Telescope (DEFT) . 
We obtained an additional 67 data points during 2016-2017. The exposure time is chosen to be 
10~mins to achieve sufficient signal-to-noise ratio (S/N $>$ 100 at 5500\AA). 
The data are then processed by an IDL-based data reduction pipeline  (Ma \& Ge, in preparation). 
This pipeline calculates the RV by matching the wavelength calibrated stellar spectra to a stellar 
template, which is generated by combining all available stellar observations of HD~26965. 
The RV data are summarized in Table~\ref{tab:tou_RV}.  

\begin{figure}
\centering
\includegraphics[width=8.2cm]{{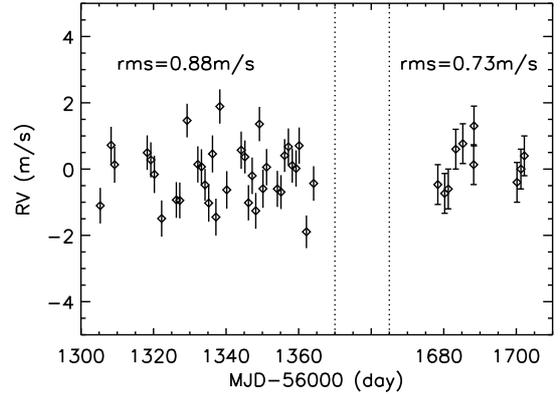}}
\caption{ Radial velocity measurements of an RV stable star, HD~10700, from 2015 to 2016. Each data 
point is obtained by combining three 10~minutes exposures of the star to remove the short-term stellar oscillation noise. 
The rms precision is between 0.7-0.9~\ms. 
\label{fig:tau}}
\end{figure}

\begin{figure}
\centering
\includegraphics[width=8.2cm]{{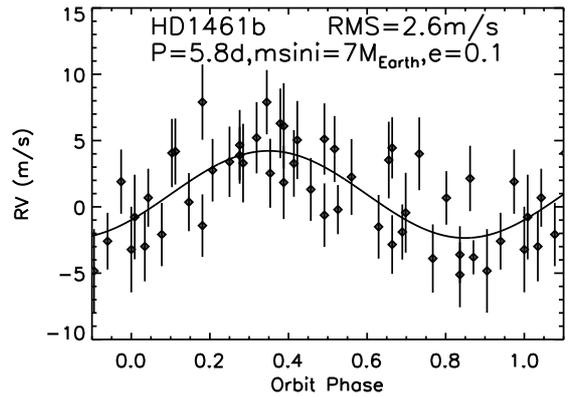}}
\caption{ Phase folded radial velocity curve of HD~1461. The RV measurements are from TOU.
The planet has a period of 5.8~d, and eccentricity of 0.1. The best-fit Keplerian orbital model is also shown, 
with a residual rms of 2.6~\ms. These RV observations demonstrate TOU has the ability to detect 
super-Earths around nearby bright stars. }
\label{fig:1461}
\end{figure}

To demonstrate the RV precision of the TOU spectrograph, we also monitored an RV stable star, HD~10700, 
and a known planet host star, HD~1461 \citep{rivera10}, using TOU between 2015 to 2016. On each night, we obtained three 
10~mins exposures of HD~10700 and combined them to calculate the RV for HD~10700 on that night. This can 
help average out the short-term periodic stellar oscillation noise \citep{dumusque11a}, thus, reducing the RV noise 
from stellar activity. The RV data for HD~10700 are displayed in Figure~\ref{fig:tau}, which shows an RV scatter of $\sim0.8$~\ms.
For HD~1461, we obtained one 30~min exposure on each night. The phased radial velocity curve is displayed in Figure~\ref{fig:1461}, 
which shows we have successfully recovered this planet RV signal.

\begin{table}
\caption{Radial Velocity Measurements of HD~26965 from TOU. The entire RV dataset is available online only. \label{tab:tou_RV}}
\begin{tabular}{ccc}
\hline\hline
FCJD & RV (\ms) & Err (\ms) \\
\hline
 2456945.795540 &  -3.6 &  1.4 \\
 2456947.791740 &  -2.2 &  1.4 \\
 2456948.844430 &   3.1 &  1.4 \\
 2456950.828580 &   0.3 &  2.2 \\
 2456951.861620 &  -4.4 &  1.5 \\
 2456952.825790 &  -3.2 &  1.3 \\
\hline
\end{tabular}
\end{table}

\subsection{Keck/HIRES RV Data }
The HIRES \citep{vogt94} spectrograph covers a wavelength range of 3700-8000\AA. It uses the iodine cell technique to measure 
radial velocities \citep{butler96}. \citet{butler17} released 20 years of precision radial velocities from HIRES on the Keck-I telescope carried out by the 
Lick-Carnegie Exoplanet Survey (LCES) Team.  HD~26965 was one of the targets covered by this survey. We found a total of 284 
observations for HD~26965 from \citet{butler17}. 
We find there exists a $\sim30$\ms offset between the RV data for 
HD~26965 before and after 2014 August. Such a big RV offset may be triggered by many causes, such as instrument effects, bad wavelength 
calibration, or data reduction pipeline glitch. Since we can not identify the exact cause for such a big RV offset, 
we decide to use only the 236 RV data points taken on 92 nights before  2014 August in this study to minimize inconsistency potentially caused by   bad RV data products. It is worth pointing out that \citet{diaz18} also only use HIRES RV 
data taken before 2014 August in their paper. The S-indices from Ca II H and K lines derived from 
Keck/HIRES data are also available from \citep{butler17}.  

\subsection{HARPS RV Data}
HARPS is a pressure and temperature stabilized spectrograph, which has a spectral resolving power of R$\sim$115,000 and 
a wavelength range between 3800 and 6300\AA \citep{mayor03}. The HARPS data for HD~26965 are downloaded 
from the ESO PHASE3 archive website. 
We find a total of 483 good exposures on 78 nights from 2003 October to 2015 January, with exposure times ranging between 30 seconds to 10~mins, 
which are labeled as HARPS old data. HARPS vacuum enclosure was opened in 2015 as part of an upgrade campaign. 
We include 82 HARPS post-upgrade RV measurements between 2015 September and 2016 March from \citep{diaz18}, which are 
labeled as HARPS new data throughtout this paper. The spectral data are reduced using 
the standard HARPS Data Reduction Software (DRS). 


\subsection{PFS RV Data}
The Carnegie Planet Finder Spectrograph (PFS) has spectral resolution of $R\sim80,000$, and is equipped with an I$_2$ cell for precise 
radial velocity measurements. Spectroscopic observations were carried out using PFS \citep{crane10} 
between 2011 and 2016. There are a total of 68 individual radial velocity measurements obtained 
on 20 different nights. The typical signal-to-noise ratio is $\sim$300 per resolution element, which 
delivers a level of $\sim$1-2~\ms RV precision. The PFS RV measurements data and corresponding 
S-indices are taken from Table~8 in \citet{diaz18}.

\subsection{CHIRON RV Data}
CHIRON is a fiber-fed high-resolution echelle spectrograph with a resolution of $R\sim95,000$ 
using the slit mode and 3$\times$1 pixel binning \citep{tokovinin13}. It has a wavelength coverage 
of 4150 to 8800\AA. The wavelength calibrated spectra are reduced using a pipeline from \citet{brewer14}.
The Doppler shifts are calculated using a standard I$_2$ technique. There are a total of 258 
measurements taken on 107 nights, with a median radial velocity error of $\sigma=1.60$\ms. 
The RV measurements are taken from Table~9 in \citet{diaz18}.

\subsection{Photometric Observations}
We acquired 1550 good photometric observations of HD~26965 during 24 
consecutive observing seasons between 1993 September and 2017 February, all 
with the T4 0.75~m automatic photoelectric telescope (APT) at Fairborn 
Observatory in the Patagonia Mountains of southern Arizona.  The T4 APT is 
one of several automated telescopes operated at Fairborn by Tennessee State 
University and is equipped with a single-channel precision photometer that 
uses an EMI 9924B bi-alkali photomultiplier tube to count photons in the 
Str\"omgren $b$ and $y$ pass bands \citep{hen1999}. Further information on the
operation of our automated telescopes, precision photometers, and observing 
and data reduction techniques can be found in \citet{h1995a}, \citet{h1995b}, 
\citet{hen1999} and \citet{ehf2003}.


\section {Stellar Parameters}
HD 26965 is the primary of a very widely separated triple system. The other two companions are an M4 dwarf and 
a white dwarf. The on-sky separation between the primary and the other 
two stars is about 82 arcsec. The estimated orbital period of this system is $\sim8000$ years \citep{heintz74}. 
This star has a star spot activity cycle period of 10.1 years \citep{baliunas95}. 

The stellar parameters are derived from excitation and ionization equilibria of Fe. 
We first normalize the stellar spectra in each order and merge them into a single spectrum in 
the spectral range 465-617~nm. We then derive equivalent widths (EWs) of Fe I and Fe II lines 
with the code TAME~\citep{kang12}, using an initial line list with 
75 Fe~I and 10 Fe~II lines from~\citet{tsantaki13}, after discarding 
Fe lines with EW~$ > 120$~m{\AA} and with EW~$ < 10$~m{\AA}. 
The stellar atmospheric parameters are computed using the code 
StePar~\citep{tabernero12}, which uses the MOOG code
\citep[in its 2014 version,][]{sneden73} and 
a grid of Kurucz ATLAS9 plane-parallel model atmospheres \citep{kurucz93}.
StePar iterates until the slopes of A(Fe~I) versus $\chi$ and A(Fe~I) versus 
log(EW/$\lambda$) are equal to zero, while imposing the ionization equilibrium 
condition A(Fe~I)$ = $A(Fe~II). StePar does a second iteration of the stellar parameter 
determination after rejecting the Fe lines with an EW and corresponding Fe abundance outliers
using a 3-$\sigma$ clipping procedure.
61 Fe~I lines and 9 Fe~II lines remain after clipping, 
which produce $T_{\rm eff} = 5072 \pm 53 $~K, $\log (g) = 4.45 \pm 0.19 $, 
[Fe/H]$ = -0.42 \pm 0.04$. Therefore, HD~26965 is a metal poor star compared to our Sun. 
Using the mass-radius-stellar parameters relation from \citet{torres10}, we calculate HD~26965 
to have a stellar mass of 0.78$M_\odot$ and a stellar radius of 0.87$R_\odot$. We listed these 
parameters in Table~\ref{tab:stellarparams}.

For comparison, the stellar parameters for HD~26965 were also derived from HARPS spectra 
by \citet{delgado17}. The abundances were determined from a standard local thermodynamic equilibrium (LTE) 
analysis using measured equivalent widths (EWs) injected into the code MOOG 
and a grid of Kurucz ATLAS9 atmospheres. 
They reported $T_{eff} = 5098 \pm 32 $~K, $\log (g) = 4.35 \pm 0.10 $,
[Fe/H]$ = -0.36 \pm 0.02$, which are consistent with our results from our TOU spectra. 
\citet{diaz18} also derived stellar parameters for HD~26965, which agree with our results as well.
Therefore, we only report  our results from TOU in Table~\ref{tab:stellarparams}. 

\begin{figure}
\includegraphics[angle=90,width=8.2cm]{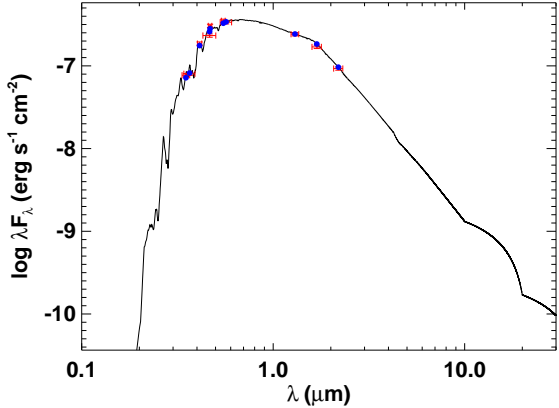}
\caption{ The observed SED from the optical through the IR for HD~26965, along with Kurucz ATLAS model 
atmosphere. Blue points represent the expected fluxes in each band based on the model, red horizontal bars are the approximate 
bandpass widths, and red vertical bars are the flux uncertainties. \label{fig:sed}
} 
\end{figure}

We also did a SED fitting using Johnson UBV \citep{ducati02}, Stromgren uvby, and JHK \citep{cutri03} band photometry, 
which is shown in Figure~\ref{fig:sed}. Using the PARSEC database of stellar evolutionary tracks 
\citep{bressan12}, we found the star has a stellar age of 6.9$\pm4.7$~Gyr. 
The error bar is big because K0 dwarfs usually stay on their main sequence for up to 15~Gyr. 

\begin{table}
\begin{center}
\caption{Stellar Parameters of HD~26965 \label{tab:stellarparams}}
\begin{tabular}{lc}
\hline\hline
Parameter & Value \\
\hline
Effective Temperature & 5072$\pm53$~K \\
log(g)    & 4.45$\pm 0.19$ \\ 
$\rm [Fe/H]$ & $ -0.42 \pm 0.04$   \\
Spectral Type & K0 V \\ 
Mass (Torres) & 0.78 $\pm$ 0.08 $M_{\odot}$\\ 
Radius (Torres) & $0.87\pm0.17 R_{\odot}$\\ 
Radius (SED) & $0.812\pm0.017 R_{\odot}$\\ 
$\rm log(R^{'}_{HK})$ & $-4.99^{a}$  \\ 
Age & 6.9$\pm$4.7~Gyr \\
$A_V$ & $0.00^{+0.01}_{-0.00}$\\
$F_{\rm bol}$   &  $5.06\pm0.12$e$^{-7}$~erg/s/cm$^2$  \\
Distance from Hipparcos & $4.985\pm0.001$ pc \\
\hline
{$^{a}$Data from \citet{jenkins11}.}
\end{tabular}
\end{center}
\end{table}

\section{MCMC Fitting of RV Data}

In order to quantify the uncertainties of the orbital parameters of the planet, we perform an MCMC analysis using the python code emcee. 
Our code follows the Bayesian method described in \citet{gregory05} and \citet{ford05, ford06}. Any noise component 
that cannot be modeled is described by a stellar jitter term $\sigma_{jitter}$ for each corresponding instrument. 

Each state in the Markov chain is described by the parameter set
\begin{equation}
\vec{\theta}=\{P_b,K_b,e_b,\omega_b,M_1,C_i,\sigma_{\rm jitter}\},
\end{equation}
where $P_b$ is orbital periods, $K_b$ is the radial velocity semi-amplitudes, $e_b$ is 
the orbital eccentricities, $\omega_b$ is the arguments of periastron, 
$M_b$ is the mean anomalies at chosen  epoch ($\tau$), $C_i$ is 
constant velocity offset between the differential RV data 
from TOU, Keck, PFS, CHIRON, and HARPS and the zero-point of the Keplerian RV model, 
and $\sigma_{\rm jitter}$ is the ``jitter'' parameter. 
The jitter parameter describes any excess noise, including both astrophysical
noise \citep[e.g. stellar oscillation and stellar spots;][]{wright05},
any instrument noise not accounted for in the quoted measurement
uncertainties, and systematic RV errors.
 
We use standard priors for each parameter \citep{gregory07}.  
The prior is uniform in the logarithm of the orbital 
period ($P_b$) from 1 to 1000 days.  
For $K_b$ and $\sigma_{\rm jitter}$ we use a modified Jefferys 
prior which takes the form of $p(x)=(x+x_o)^{-1}[\log(1+x_{max}/x_o]^{-1}$ 
, where $x_o=0.1$~${\rm m \, s^{-1}}$ and $x_{\mathrm
max} = 20$~${\rm m \, s^{-1}}$ \citep{gregory05}.  
Prior for $e_b$ is uniform between zero and unity. 
Priors for $\omega_b$ and $M_b$  are uniform 
between zero and $2\pi$.
For $C_i$, the prior is uniform between min($v_i$)-50\ms and  max($v_i$)+50\ms, 
where $v_i$ are the set of radial velocities obtained from each of the RV instruments. 
We verified that the chains did not approach the limiting values of $P_b$, $K_b$, 
and $\sigma_{\rm jitter}$. 

Following \citet{ford06}, we adopt a likelihood (i.e., conditional probability of 
making the specified measurements given a particular set of model parameters) of
\begin{equation}
p(v|\vec{\theta},M) \propto \prod_k \frac{\exp[-(v_{k,\theta}-v_k)^2/2(\sigma^2_{k,obs}+\sigma^2_{\rm jitter})]}{\sqrt{{\sigma_{k,obs}}^2+{\sigma_{\rm jitter}}^2}},
\end{equation}
where $v_k$ is the observed radial velocity at time $t_k$, $v_{k,\theta}$ is the
model velocity at time $t_k$ given the model parameters
$\vec{\theta}$, and $\sigma_{k,obs}$ is the measurement uncertainty
for the radial velocity observation at time $t_k$.

\begin{figure}
\centering
\includegraphics[width=8.2cm]{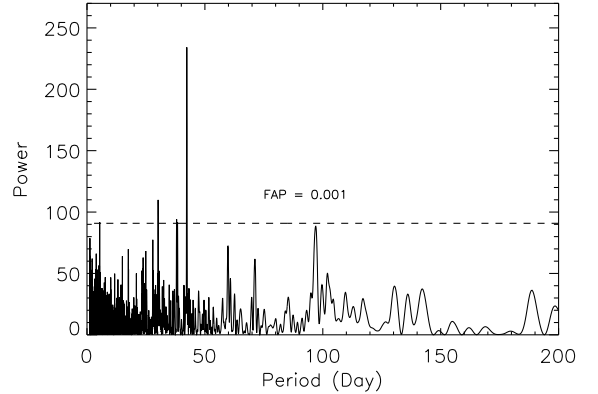}
\caption{ Generalized Lomb-Scargle periodogram for the RV data, which clearly shows the peak around 42 day. The horizontal dashed line 
shows the 0.1$\%$ significance level determined using 10000 bootstrap resamplings and a search window of [2, 200]~days.}
\label{fig:mcmc1}
\end{figure}

\begin{figure*}
\centering
\includegraphics[width=14cm]{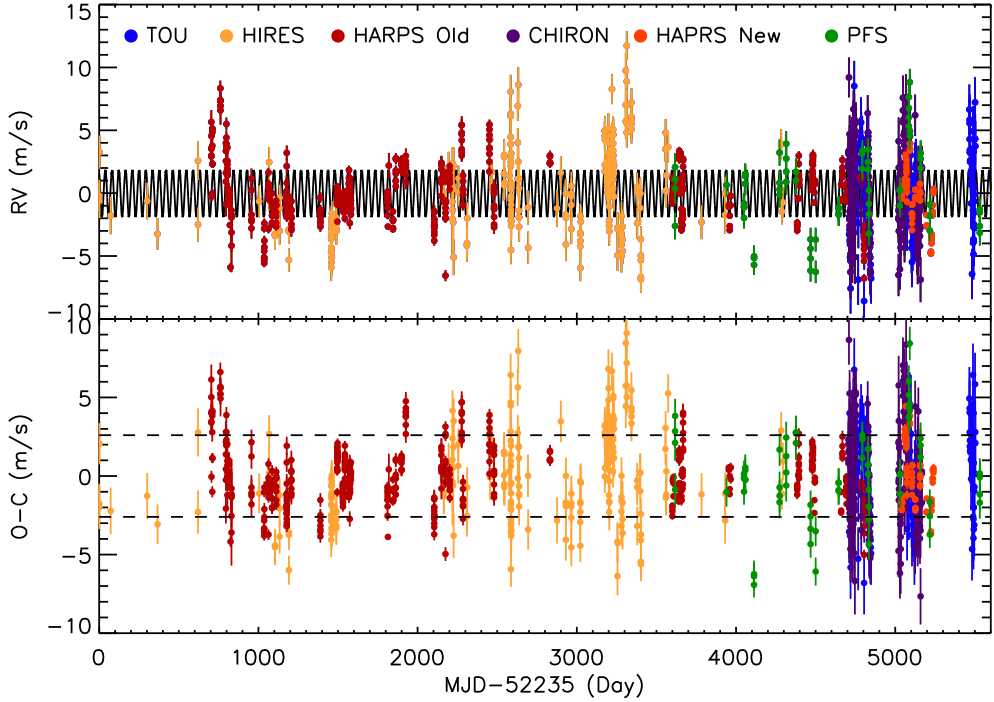}
\caption{ Best-fit 1-planet Keplerian orbital model for HD~26965. The RV data points from Keck/HIRES, HAPRS, PFS, CHIRON, and TOU are 
shown as yellow, red, and blue dots.
The black solid line is the maximum likelihood model, with the orbital parameters listed in Table \ref{tab:params}.
The bottom panel shows the RV residuals after subtracting the best-fit Keplerian RV model. The horizontal dashed liens show the range of the rms value, $\pm2.6$~\ms. 
}
\label{fig:mcmc}
\end{figure*}

\begin{figure}
\centering
\includegraphics[width=8.2cm]{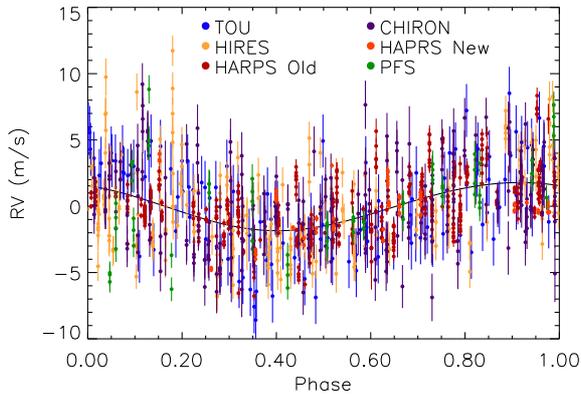}
\caption{ Phase-folded RV curve of HD~26965.
The black solid line is best-fit 1-planet Keplerian orbital model, with the orbital parameters listed in Table \ref{tab:params}.
}
\label{fig:mcmc2}
\end{figure}

In Table~\ref{tab:params} we show the final parameters and uncertainties obtained with our MCMC analysis. 
We performed a simultaneous fit of the planetary signal and the activity induced RV jitters using the TOU, 
Keck, PFS, CHIRON, and HARPS data. From the periodogram of the RV data shown in Figure~\ref{fig:mcmc1}, we find a strong periodic signal around 42-day.  
The $0.1\%$ significance level shown on this plot is determined after 10000 bootstrap resampling. 
In Figure~\ref{fig:mcmc} we show the best Keplerian orbital fit to the RV data which is attributed to the planet candidate HD~26965b. The phase-folded 
RV curve is shown in Figure~\ref{fig:mcmc2}. 
The RMS of the residuals is 2.6\ms, and we did not find any additional peak with significance above $0.1\%$ from the periodogram of 
RV residuals. 
Adopting a stellar mass of 0.78$M_\odot$, we derived the minimum mass of the planet $m\sin i=9.7\pm1.3M_{\earth}$.
%

\begin{table}
\caption{MCMC Posteriors for the Keplerian Orbital Fitting}
\begin{tabular}{cccc}
\hline\hline
Parameter & Credible Interval & Maximum Likelihood & Units \\
\hline
$P_{b}$ & 42.378 $\pm 0.01$ & 42.378 & days\\
 $T\rm{conj}_{b}$ & 5886.76 $^{+0.62}_{-0.58}$ & 5886.50 & JD\\ 
$e_{b}$ & 0.04$^{+0.05}_{-0.03}$ & 0.02 & \\
$\omega_{b}$ & 2.6 $^{+2.2}_{-1.4}$ & 2.2 & radians\\
$K_{b}$ & 1.81 $^{+0.10}_{-0.10}$ & 1.80 & m s$^{-1}$\\
$C_{\rm TOU}$   & -0.32 $\pm 0.26$ & -0.31 & m s$^{-1}$\\
$C_{\rm HIRES}$ & 0.23 $\pm 0.23$ &  0.24 & m s$^{-1}$\\
$C_{\rm PFS}$ & -0.37 $\pm 0.39$ &  -0.38 & m s$^{-1}$\\
$C_{\rm CHIRON}$ & 0.16 $\pm 0.19$ &  0.16 & m s$^{-1}$\\
$C_{\rm HARPS_old}$ & 0.05 $\pm 0.09$ & 0.0.04 & m s$^{-1}$\\
$C_{\rm HARPS_new}$ & -0.02 $\pm 0.23$ & -0.02 & m s$^{-1}$\\
$\sigma_{\rm TOU}$ & 2.06 $^{+0.28}_{-0.27}$ & 2.03 & $\rm m\ s^{-1}$\\
$\sigma_{\rm HIRES}$ & 3.08 $^{+0.19}_{-0.17}$ & 3.06 & $\rm m\ s^{-1}$\\
$\sigma_{\rm PFS}$ & 2.82 $^{+0.32}_{-0.28}$ & 2.70 & $\rm m\ s^{-1}$\\
$\sigma_{\rm CHIRON}$ & 2.41 $^{+0.18}_{-0.16}$ & 2.39 & $\rm m\ s^{-1}$\\
$\sigma_{\rm HARPS_old}$ & 1.81 $^{+0.07}_{-0.6}$ & 1.80 & $\rm m\ s^{-1}$\\
$\sigma_{\rm HARPS_new}$ & 1.91 $^{+0.19}_{-0.15}$ & 1.88 & $\rm m\ s^{-1}$\\
$m\sin i$ & 8.47$\pm0.47$ & 8.43 & $M_{\earth}$ \\
\hline
\end{tabular}
\label{tab:params}
\end{table}

\section{Planet, or Stellar Activity? \label{sec:activity}} 
In this section, we discuss the possibility that the RV signal is actually produced by stellar rotation modulated activity \citep[like starspots, plagues, and convection inhibition; ][]{dumusque11a}. We first determine the possible stellar rotation periods from both stellar activity index and photometric data, and compare them with the period of the RV signal. We then assess the stability of this 42-day RV signal by investigating the evolution of this 42-day RV signal against the number of RV observations. We also discuss the possible RV jitter induced by stellar surface activity and compare it with the amplitude of the 42-day RV signal. Line bisector analysis is conducted in the end to investigate the impact of stellar activity on RV measurements. All of the studies except the activity period in this section support the planet origin of the 42-day RV signal, which are summarized in Table~\ref{tab:discuss}.

\begin{table*}
\caption{Summary of evidence supporting planet origin or stellar activity origin of the detected 42~d RV signal.}
\begin{tabular}{cc}
\hline\hline
Evidence & Planet or Activity Origin \\
\hline
$P_{\rm rotation}$ close to 42~d &  Activity  \\
Sharp 42~d peak from RV periodogram & Planet \\ 
No clear 42~d peak from $S_{\rm MW}$ periodogram &  Planet  \\
Clear 42~d peak from magnetic quiet phase RV periodogram &  Planet  \\
Clear 42~d peak from magnetic active phase RV periodogram & Planet  \\
Strong 39~d signal from $S_{\rm MW}$ in magnetic quiet phase & Planet  \\
Strong 41~d signal from $S_{\rm MW}$ in magnetic active phase & Planet   \\
No 42~d peak from 23 years high precision photometric campaign  & Planet \\
Star spots $K_{\rm spot}<0.3$\ms from simulation & Planet \\ 
Inhibition of convection $K_{\rm inh}<0.3$\ms from linear interpolation & Planet  \\
Active phase and quiet phase have similar $K$ values detected & Planet \\ 
No strong correlation between RV and BIS & Planet \\ 
\hline
\end{tabular}
\label{tab:discuss}
\end{table*}

\subsection{Stellar Activity Index and Rotation Period} 
\begin{figure}
\includegraphics[angle=0,width=8.2cm]{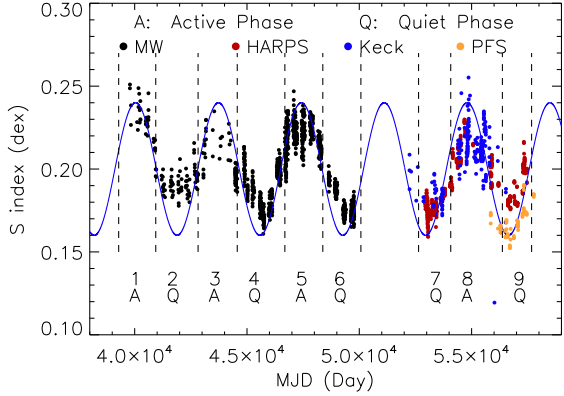}
\caption{ Measurements of Ca II HK S index of HD~26965. The data are collected from Mount Wilson observatory (black), HARPS (red), Keck (blue), and PFS (orange). We over-plotted a Sine curve with a 10.1~yr period to show the sun-like magnetic cycle. The data are separated into alternative magnetic quiet and active phases, which are marked by a zone number from 1 to 9. In the plot, `A' represents magnetic active phase, and `Q' represents magnetic quiet phase.  
\label{fig:HK} } 
\end{figure}

We first determine the rotation period of the star using the stellar activity S index. The S index, calculated from the singly ionized calcium H \& K line core emission flux \citep{wilson78}, is the most commonly used index of stellar magnetic activity. To put constraints on the rotation period and the magnetic activity cycle of the star, we re-examined the Ca II HK S index data from Mount Wilson ($S_{\rm MW}$), HARPS, Keck, and PFS. Olin Wilson's HK Project at the Mount Wilson Observatory (MWO) regularly observed the Ca II HK emission for a sample of over 100 bright dwarf stars beginning in 1966 to characterize magnetic variability of stars other than the Sun \citep{wilson78}. We obtained the Mount Wilson data from the National Solar Observatory (NSO) and present them in Figure~\ref{fig:HK}. From the plot, we can clearly see a 10.1~yr magnetic cycle \citep[also reported in][]{baliunas96}, which shows the star periodically enters into an active phase after a relatively quiet phase. Since there is a $\sim10$~yr magnetic cycle, we need to remove its signal first before we can examine the short-term $S_{\rm MW}$ variations caused by stellar rotation. We used a spline function with a breakpoint of 200 days to de-trend the $S_{\rm MW}$ values. For the S index data from HAPRS, KECK, and PFS, we use a breakpoint of 500 days because there are not as many data points available as there are from Mount Wilson. After examining the periodogram for the de-trended $S_{\rm MW}$, we do not find a clear peak at 42.4~d. From the study of the Sun, we learn that the active regions are not always concentrated on certain longitudes, hindering the detection of its stellar rotation signal during the active phase. Therefore, we decide to separate these S index data alternatively into the quiet phase and active phase as shown in zone 1 to 9 in Figure~\ref{fig:HK} to check for occasionally strong periodic signals.



\begin{figure}
\includegraphics[angle=0,width=8.2cm]{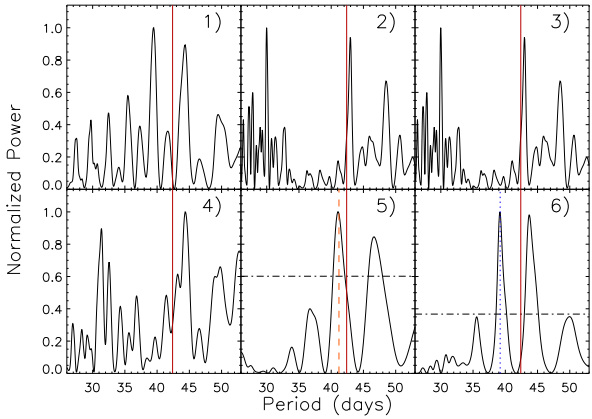}
\caption{ Periodograms for the Mount Wilson Ca II HK S index. 
From top left to bottom right, the panels show periodograms of $S_{\rm MW}$ from zone 1 to zone 6 marked as in Figure~\ref{fig:HK}. 
We searched for strong periodic signals around 
42~days in each zone. The horizontal dashed lines shown in panel 5 and 6 mark the $0.1\%$ significance level, which is 
derived using 10000 times bootstrap resampling and a search window of [2, 300]~days.  The red vertical solid line in each panel 
shows the period of the planet candidate HD~26965b at 42.4~d. The orange vertical dashed line in panel 5 
shows the period of a strong signal in a magnetic active phase at 41.2d. 
The blue dotted vertical  line in panel 6 shows the period of the strong signal in a magnetic quiet phase at 39.2d. 
\label{fig:period_active_quiet_s} } 
\end{figure}

\begin{figure}
\includegraphics[angle=0,width=8.2cm]{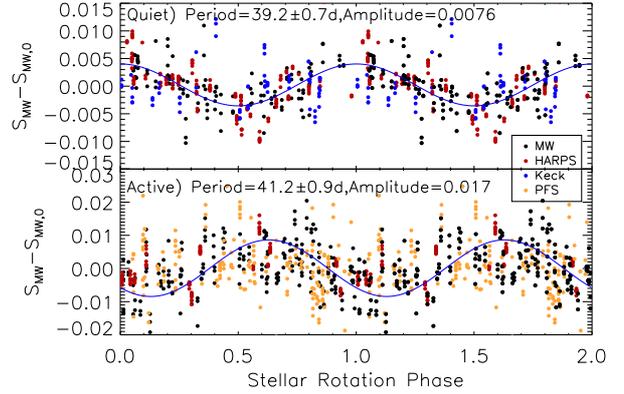}
\caption{ Phase folded Mount Wilson, HAPRS, Keck, and PFS S index measurements of HD~26965. 
The data shown in the top panel are from the magnetic quiet phase, and the data in the bottom panel are from the 
magnetic active phase. We fitted a Sine function to the S index after subtracting their median value, and 
calculated the best-fit period and amplitude for both phases using the Mount Wilson data. The S indices from HARPS, 
Keck, and PFS were not used in the Sine curve fitting, but they are consistent with the rotation period fitted 
from the Mount Wilson data. The active phase shows a bigger S index variation amplitude due to a 
more active status of the star compared to the quiet phase.  
\label{fig:HK2} } 
\end{figure}

\begin{figure}
\includegraphics[angle=0,width=8.2cm]{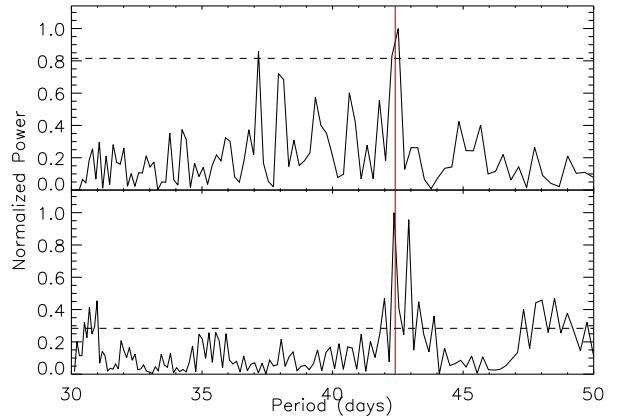}
\caption{ Periodograms for the radial velocities for alternating active (the top panel) and quiet phases (the bottom panel), respectively. 
Both periodograms show a clear peak around 42.4~d, which is marked with a vertical red solid line. 
The horizontal dashed line in each panel marks the $1\%$ significance level, which is derived using a 1000 times bootstrap 
resampling and a search window of [2, 300]~days.  
\label{fig:period_active_quiet_rv} } 
\end{figure}

The periodogram for these marked zones are shown in Figure~\ref{fig:period_active_quiet_s}. We found strong signals in zone 5 and zone 6, 
corresponding to an active phase and a quiet phase, respectively. Since the S index data from HAPRS, KECK, and PFS do not have error bars, we decided to not run 
periodograms and Sine curve fitting on these data. The S index data from Zone 5 and Zone 6 together with their best Sine curve fit are displayed in Figure~\ref{fig:HK2}. HAPRS, Keck, and PFS S index data from Zone 8 and 9 are also displayed in Figure~\ref{fig:HK2}, which support consistent periodic signals found from the Mount Wilson data in both the active and quiet phases. The resulting best-fit periods for the active and quiet phases are $41.2\pm0.9$~days and $39.2\pm0.7$~days, respectively, with errors estimated from the FWHMs of the peaks in the periodograms. The period variation can be explained by the differential rotation of the star and different locations of active regions similar to our Sun. For comparison, the periodograms of radial velocities in both the active phase and quiet phase (Figure~\ref{fig:period_active_quiet_rv}) clearly show a stable peak around 42.4~d. The fact that periodograms for the activity indicator $S_{\rm MW}$ show multi-peaks, 
and none of these strong peaks sit exactly at 42.4~d, is an important piece of evidence supporting the planet origin of this 42.4~d signal.
The best-fit amplitudes of the S index modulation (Figure~\ref{fig:HK2}) are 0.008 and 0.017 in the quiet phase and the active phase, 
which demonstrates there is stronger stellar surface activity during the magnetic active phase than during the quiet phase. 
As explained later in section~\ref{section:activity}, this finding also supports the planet origin of this 42.4~d signal. 

There are several other estimates of the stellar rotation period from previous work, like 43~days using 25~years of the Ca II HK index measurements from Mount Wilson Observatory \citep{baliunas96},  42.2~days using the calibrated stellar activity-rotation-age relation \citep{mamajek08, lovis11}, 43.7~days using the relation from \citep{mascareno15, mascareno16}. All of the rotation periods estimated above, including our own result, are close to the period of the RV signal detected in this paper, which warrants further discussion of the possibility that this RV signal is induced by stellar surface magnetic activity. 
 

\subsection{Photometric Results and Rotation Period}

\begin{figure}
\includegraphics[angle=0,width=8.2cm]{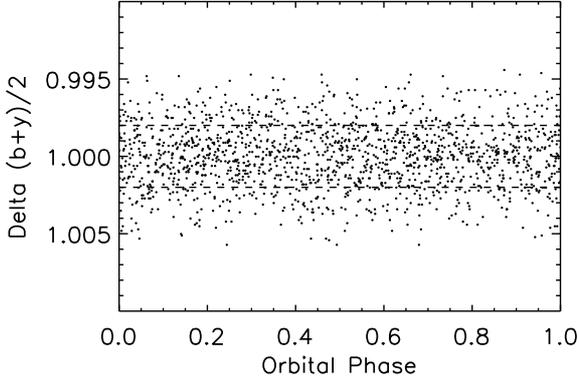}
\caption{ Photometric data of HD~26965 acquired with the T4 APT at Fairborn Observatory from 1993 to 2017.  A total of 1550 differential photometric data points of HD~26965 are phased to the period of the planet RV signal. The two horizontal dashed line correspond to the $\pm1\sigma$ range of the brightness level of the 1550 observations.  \label{fig:photo}
} 
\end{figure}

\begin{figure}
\includegraphics[angle=0,width=8.2cm]{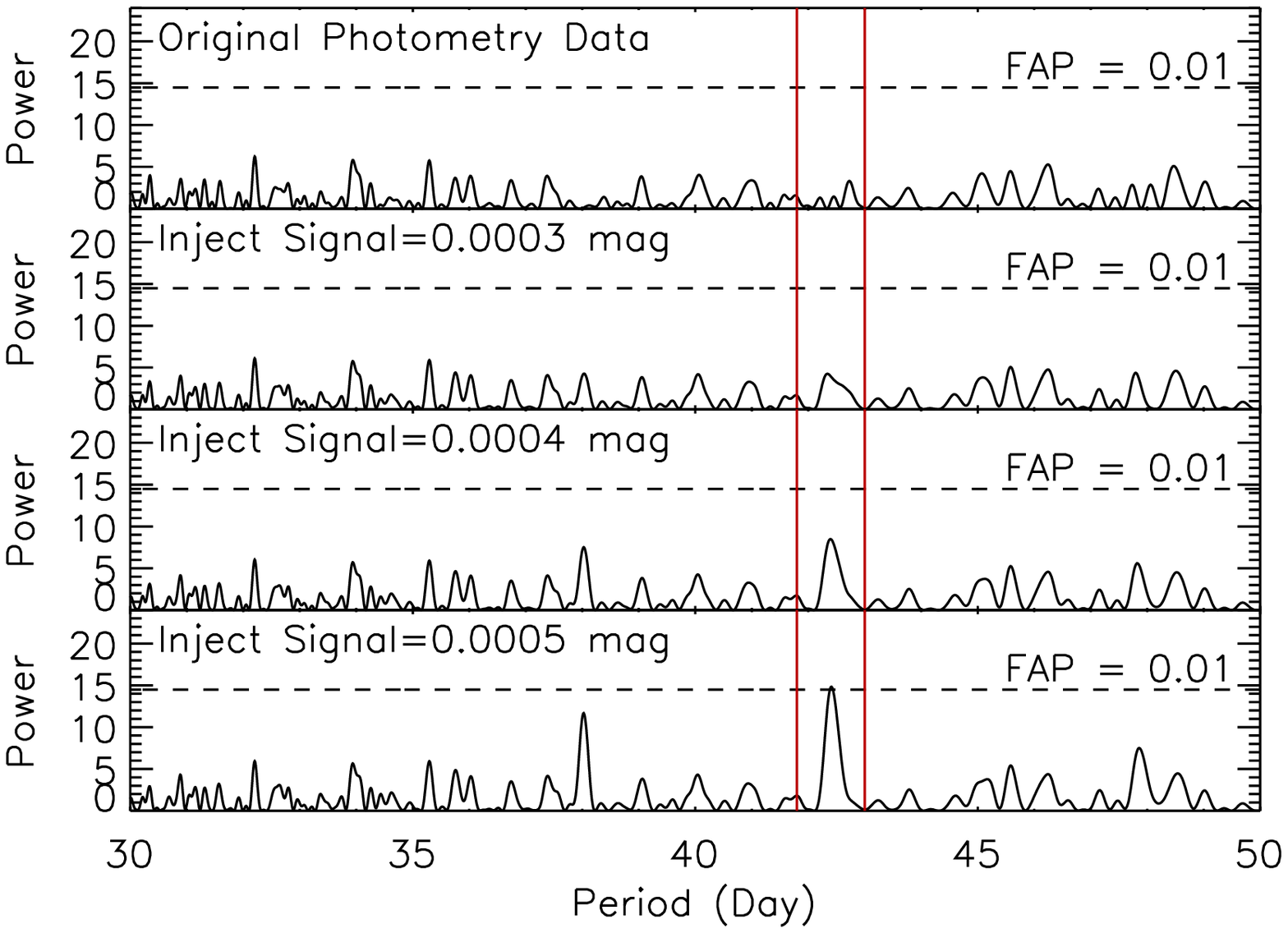}
\caption{ Periodogram of photometric data for HD~26965 from Fairborn. There is no significant peak except the 1-day alias in the top panel. The double red solid lines mark the region around $P=42.38$~d, which is the planet orbital period. In the second, third and fourth panels, we inject a periodic signal at semi-amplitudes of 0.0003, 0.0004, and 0.0005~mag to the original photometric data. The peak at 42~days becomes higher as the injected signal becomes stronger. Comparing the simulated periodogram with the original periodogram in the top panel, we can rule out a periodic signal stronger than $0.0004$~mag. We note here that the 1 year alias of the 42.4~d signal, which sits at 38~d, also becomes stronger. \label{fig:photo_period} } 
\end{figure}

In this section we search for a periodic rotation signal from photometric data. In Figure~\ref{fig:photo} we plot all 1550 photometric measurements from Fairborn Observatory plotted against the orbital phase of HD~26965b with $P_b=42.38$~d. The standard deviation of these data from their mean is 0.002~mag. A least-squares Sine-curve fit to the phased data gives a full amplitude of $0.0000\pm0.0002$~mag. We can rule out a sinusoidal brightness variation larger than 0.0006~mag at 3-$\sigma$ confidence at the orbital period of the planet candidate. To put further constraint on the upper limit of a detectable periodic photometric signal that might be induced by starspots, we did a simple simulation by adding a sinusoidal photometric signal to the photometric data collected from Fairborn. The period for this signal is set to be the same as the period of the planet signal ($P=42.38$~d), and the semi-amplitude is chosen to be $0.0003$, $0.0004$, and $0.0005$~mag. From the periodogram shown in Figure~\ref{fig:photo_period}, we cannot detect the $0.0002$~mag signal, but can start to detect this periodic photometric signal at $0.0005$~mag. This simulation helps rule out a detectable periodic photometric signal with a semi-amplitude larger than 0.0004~mag. We use this information to put a constraint on stellar RV jitters induced by starspot activity in the next section.

%
%

\begin{figure}
\includegraphics[angle=0,width=8.2cm]{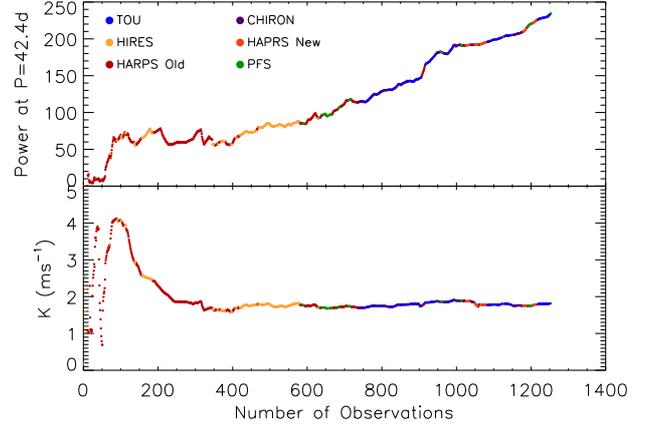}
\caption{ Evolution of the significance of the detected signal and its semi-amplitude as a function of the number of measurements. 
The semi-amplitude is derived using a fixed orbital period and eccentricity. 
\label{fig:evolution1} } 
\end{figure}

\begin{figure}
\includegraphics[angle=0,width=8.2cm]{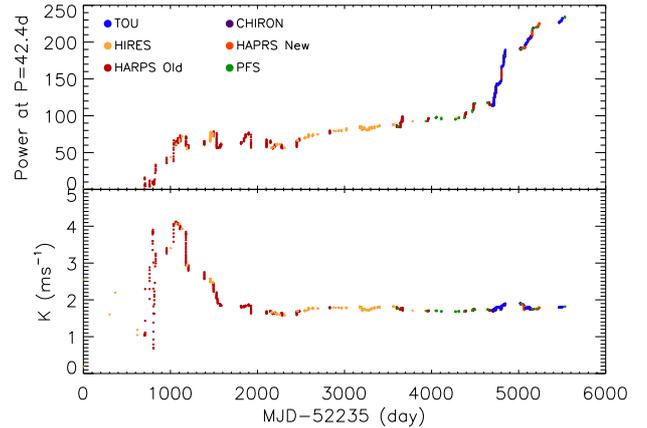}
\caption{ Evolution of the  significance of the detected signal and its semi-amplitude as a function of the date of the observations.
The semi-amplitude is derived using a fixed orbital period and eccentricity. 
 \label{fig:evolution2} } 
\end{figure}

\subsection{RV Signal Evolution \label{section:signal_evolution}}
To assess the stability of this RV signal, we study the evolution of this 42-day RV signal against the number of observations and date of observations 
in this section. The significance power of this 42-day signal is calculated using a generalized Lomb-Scargle periodogram (GLS) following \citet{zechmeister09}. Figure~\ref{fig:evolution1} shows the evolution of the signal significance and its velocity semi-amplitude  ($K$) against 
number of observations. The evolution of this 42-day signal is steady and continuous after $N=400$ data points. Figure~\ref{fig:evolution2} shows the 
evolution of  the signal significance and its velocity semi-amplitude ($K$)  against the date of observation. 
By studying the evolution against the date of observations, we can also
see the importance of the high-cadence Dharma survey observations. The slope of the significance increased 
after the start of the TOU nightly cadence campaigns in 2015, which means that the time needed for the detection of short-period planets 
decreases significantly, similar to what happened with the discovery of Proxima b \citep{ang16}.
If this 42-day RV signal is generated by stellar activity, it should not be so stable given the large amplitude variation of stellar activity strength 
from its magnetic cycle. These two plots support the stableness of the orbital parameters and, thus, the planetary source of this 42-day RV signal \citep{mascareno17}. 

\begin{figure}
\includegraphics[angle=0,width=8.2cm]{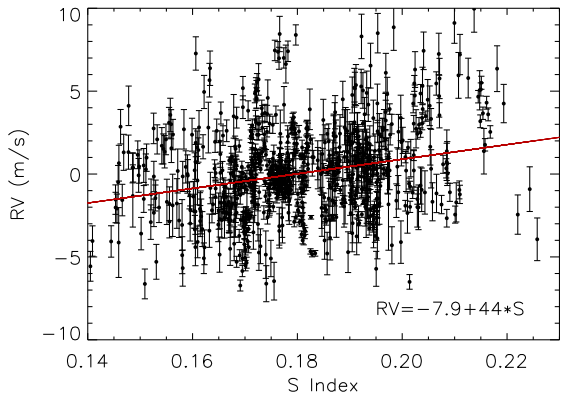}
\caption{ Correlation between Ca II HK S index and radial velocity measured by HARPS, HIRES, and PFS. 
The red solid line shows the best linear-fitting result ($RV=-7.9+44\times S$). 
 \label{fig:HK_RV} } 
\end{figure}

\subsection{Stellar Activity RV Jitter \label{section:activity}}
In this section, we provide additional evidence to support the planet nature of this RV signal. There are several stellar 
activity sources that can generate an RV signal detected in this paper, like dark spots, plages, and convection inhibition \citep{vanderburg16}. 
We discuss these possibilities in this section. First, we simulated the RV signal induced by dark spots using a modified code of SOAP2.0 
(Spot Oscillation And Planet, \citealt{dumusque14}; Kimock et al., in preparation). By setting a photometric variation semi-amplitude of  
$0.0004$~mag, the upper limit derived from the last section, and using a single spot model for simplicity, the maximum 
RV signal generated has $K_{\rm spot}<0.3$~\ms, which is a factor of seven times less than the RV signal detected. We also explored a more realistic multi-spot 
model similar to the Sun in our simulation; the RV signal generated is even smaller in the $K_{\rm spot}$ value, similar to what was concluded in 
the previous study by \citet{dumusque14}. 

Secondly, the inhibition of convection due to surface magnetic activity can generate artificial redshift. Solar-like magnetic cycles are 
characterized by an increasing filling factor of active regions when the activity level increases. Because convection is strongly 
reduced in active regions as a result of the magnetic field, the star appears redder (thus positive radial velocity) during its high-activity
phase. A positive correlation between the RVs and the activity level is therefore observed \citep{meunier10}. However, 
neither \citet{santos10} nor \citet{lovis11} found any significant RV variation from HD~26965 due to convection 
inhibition during its magnetic cycle ($<2$\ms). Here we reproduced this correlation in Figure~\ref{fig:HK_RV} using 
the RV and Ca II HK S-indices data from HARPS, PFS, and Keck/HIRES, and made a linear fit to this correlation. 
We came to the same conclusion as stated by \citet{santos10}, that there is a 
very weak correlation between the RV and Ca II HK S-index. Using the linear relation from Figure~\ref{fig:HK_RV},  the convection 
inhibition can generate an RV variation with a semi-amplitude $K_{\rm inh}\sim0.2$\ms within a 42~day period when the periodic coherent 
Ca II HK S-index variation is as small as $\sim$0.004 (semi-amplitude) shown in the top panel of Figure~\ref{fig:HK2}. Clearly this is not big enough 
to explain the semi-amplitude of $\sim$1.8\ms RV signal detected.  

\begin{figure}
\includegraphics[angle=0,width=8.2cm]{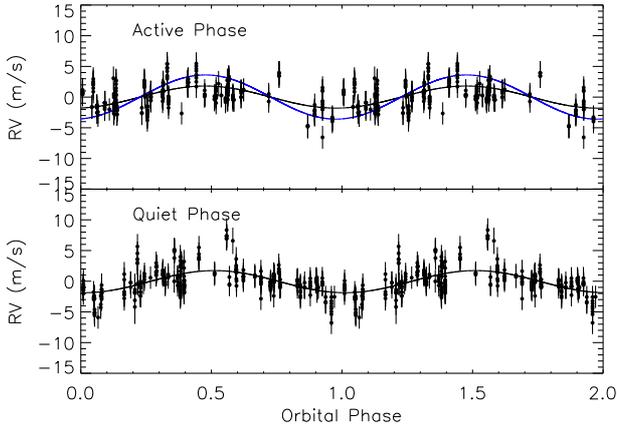}
\caption{ Radial Velocity data from HARPS, which are divided into quiet (bottom panel) and active (top panel) 
phases according to the magnetic cycle of the star. We use $P=42.38$~d and $e=0$ to fit the RV 
data to study the RV amplitude variation from the quiet phase to active phase. 
The black lines in both panels show the best-fit RV models, and the blue solid line in the 
top panel shows a model with an RV amplitude two times of that from the quiet phase. 
The fitting result does not support the hypothesis that the RV amplitude in the active phase became 
twice as large as that in the quiet phase, which supports the planet origin of the 42-day RV signal. 
\label{fig:HARPS_active_quiet_rv}
} 
\end{figure}

The above discussions show that the most common types of magnetic activity cannot induce the RV signal detected in this investigation. 
Next we present additional evidence against the activity origin of this 42-day RV signal. In the previous section, we identified the magnetic active phases and quiet phases based on over 30~years of Ca II HK index measurements of HD~26965 (Figure~\ref{fig:HK}). We then divided all the RV measurements into either an active phase or a quiet phase. The phased curves of Ca II HK index measurements in Figure~\ref{fig:HK2} show that the surface activity filling factor varies a factor of two between the quiet phase and the active phase. 
\citet{lanza16} found there is a positive correlation between solar RV variation and the level of chromospheric activity measured using Ca II HK index. 
If the coherent 42-day RV signal detected is from stellar activity modulated by the rotation, we expect the RV amplitude would become two times larger in the active phase 
than in the quiet phase because the coherent S-index variation in the active phase is twice as large as that in the quiet phase. Since HARPS RV data span over more than 10 years, we choose to use HARPS RV data to test this scenario. After dividing HAPRS RV data into active phase and quiet phase data, we did a Keplerian RV fitting for each phase. During the fitting, we fixed two parameters with $P=42.38$~day and $e=0.0$ to focus on the velocity semi-amplitude variation. The fitting results are shown in Figure~\ref{fig:HARPS_active_quiet_rv}, with $K_{\rm active} =1.7\pm0.3$\ms and $K_{\rm quiet} = 1.8\pm0.4$\ms. 
The fact that RV amplitude does not increase significantly from the quiet phase to the active phase again supports the planet origin of this 42-day 
RV signal. 

In \citet{mascareo17b}, they studied the radial velocity signal induced by stellar activity and rotation among 55 late-type dwarf stars using HARPS data. 
They derived an empirical relationship between the mean level of chromospheric emission and the radial velocity semi-amplitude shown in 
Figure~9 of their paper. Using $\rm log(R^{'}_{HK})=-4.99$ from \citet{jenkins11}, we estimate that the expected RV signal at this level of stellar activity 
is $\sim$0.35\ms, which is much smaller than the stable $\sim$1.8\ms RV signal detected. This also supports the planet origin of this 42-day RV signal.

\citet{vanderburg16} conducted extensive sets of simulations to study the impact of stellar surface activity modulated by stellar rotation 
on the ability to detect planets using the radial velocity technique. In their simulations, they usually assume RV observations with ideal sampling, 
which is similar to the situation in this paper when combining the Keck/HIRES, HARPS, PFS, CHIRON, and TOU observations. 
They found that an activity signal identified from the RV periodogram at a period of $P_{\rm activity}$ usually has a width as large as 
$\sim0.1*P_{\rm activity}$ because of the lifetime of spots and differential stellar rotation. 
While in our RV periodogram, the peak at 42~day is very sharp with a $FWHM=0.3$~d. This points to the
 likely conclusion that the coherent 42-d RV signal found from HD~26965 is induced by a planet, not magnetic activity.

\subsection{Line Bisector Analysis}
Line bisector analysis is another diagnostic tool to investigate the possible impact of stellar activity on RV measurements \citep{toner88, queloz01, wright13}. 
Here we performed a correlation study of the Bisector Inverse Slope (BIS) and 
the radial velocities from the fiber-fed TOU spectrograph. 
In our RV data pipeline, we do not use the Cross-Correlation Function (CCF) method employed by HARPS  (Ma \& Ge, in preparation). Instead we use a template matching method similar to \citet{zechmeister18}. Thus, we do not have a traditional CCF product 
from our pipeline to calculate the BIS. Instead, we used the reduced-$\chi^2$ function from our template matching pipeline to calculate the BIS. 
Similar to the method employed by \citet{santos02} on their analysis of the HARPS CCF, we compute the bisector 
velocity for 10 different levels on the reduced-$\chi^2$ function from TOU. The Bisector Inverse Slope is calculated by averaging 
the upper and lower bisector points before subtracting one from the other. Here we choose the definition from \citep{queloz01} 
where they use the 10-40$\%$ and 55-85$\%$ CCF depth \citep[see also][]{wright13}. 
This quantity is equivalent to the BIS from the CCF method since both trace the asymmetry in the absorption line 
profile variation caused by stellar surface activity. 
The Spearman's rank correlation coefficient between the RV and the BIS 
is 0.15 with a significance level of 0.08, which suggests there does not exist a strong correlation between the RV and BIS. 
This is more strong evidence to argue against the activity origin of the 42-day RV signal.


\section{Conclusion and Discussion}
In a search through the early RV data from the DPS survey, we discovered an RV signal consistent with a super-Earth 
orbiting a V=4.4 K dwarf, HD~26965. Additional RV data were found from the Keck archive and HARPS archive. 
After combining the RV data from the DPS survey with data from Keck/HIRES and HAPRS, we use an MCMC code to 
find the best-fit orbital parameters with an orbital period of 42.38$\pm0.01$~d, eccentricity of 0.04$^{+0.05}_{-0.03}$, and 
velocity semi-amplitude of $1.81\pm0.10$\ms. Adopting a stellar mass of 0.78$M_\odot$ for HD~26965, the minimum 
mass for the planet is 8.47$\pm0.47$ Earth masses, which puts it in the super-Earth mass range. 

The 42-day period RV signal has also been reported in \citet{diaz18}. We privately communicated about 
our discoveries during the 2017 summer Extremely Precise RV meeting (EPRV) at the Pennsylvania State 
University. The best Keplerian solution reported from their modeling 
has a $P=42.364\pm0.015$~d, $e=0.017\pm0.046$, and $K=1.59\pm0.15$\ms. The periods reported from our paper and 
their paper are similar. The fact they reported a smaller RV signal may be related to their 
modeling of red noise and linear correlations with stellar activity indicators, in addition to the white noise used in our RV modeling. 

The conclusion from \citet{diaz18} is that the RV signal can be either from a planet or from stellar activity. 
Using Ca II HK index variation, we find this star does show a long-term magnetic cycle of $\sim10.1$~yr. 
The fact that the orbital period of the planet is close to the rotation period of the star is concerning, 
because stellar rotation modulation of magnetic activity can mimic planet signals \citep{saar97, desort07, ma12}. 
For instance, \citet{queloz01} found that the RV variation of a G0V star, HD 166435, is from surface spot activity, 
not a planet. \citet{huel08} and \citet{huerta08} also found that two previously claimed exoplanets are actually 
caused by starspots on the stellar surface. \citet{mahmud11} showed that cool surface spots could 
cause the periodic RV variability on a T Tauri star. 

By carefully examining the RV data in the active phase and quiet phase of the star, and after carefully 
considering all possible stellar activity sources, we concluded that the coherent signal seen from HD~26965 is most likely
 from a planet, with some RV noise contributed by stellar activity. In addition, the sharpness of the 42~day 
 peak in the RV periodogram also supports the planet origin of the 42-d RV signal as 
active regions on the stellar surface modulated by differential rotation of the star normally 
reveal themselves as a group of peaks around the mean rotation period \citep{vanderburg16}.  
The evolution of the RV semi-amplitude to a stable value  after several years of observations provides 
additional strong support for the planet origin of this 42-day RV signal. 
Our high quality photometric dataset helps rule out any significant photometric variation at 42~days. 
This also supports the planet origin of this 42-day RV signal. This plethora of evidence allows us to draw the 
conclusion that this 42-day RV signal is from a planet, unlike the uncertainties reported in \citet{diaz18}.  

Currently there are several planet systems known 
to host planets with periods close to the rotation period of the star \citep{dragomir12, haywood14}, 
or close to the magnetic cycle period of the star \citep{wright08, kane16}. For example, 
\citet{dragomir12} found a Jupiter-mass planet orbiting HD~192263 with a period of 
24.4~days using radial velocity from Keck/HIRES and CORALIE, and derived a stellar 
rotation period of 23.4~days using photometric data. We note here that \citet{henry02} initially suggested that
 the RV signal around HD~192263 is caused by rotational modulation of surface activity. 
But \citet{santos03} used significant changes of photometric patterns over time and 
3 years of coherent RV observations to prove that the signal is indeed from a planet. In the case of HD~26965, we cannot 
derive a precise stellar rotation period because our ground-based photometric data did not 
reveal any significant periodic variation around 42~days. Future high precision space photometric observations 
can help detect possible small photometric variation from the star ($< 400$~ppm) caused by the surface spots.
As pointed out by \citet{vanderburg16}, it is very important for next generation RV planet surveys 
to have simultaneous photometric observations for measuring rotation periods and activity signals. 
Our data analysis also demonstrates the importance of having simultaneous photometry for the purpose of 
disentangling planet signals from magnetic activity signals.


HD~26965 is a very bright metal poor star with V=4.4. This makes it the second brightest star in the night sky 
with a super-Earth detection so far, just behind HD~20794 \citep[V=4.3][]{pepe11}. One interesting 
fact is that HD~20794 has a similar metallically (Fe/H$=-0.4\pm0.1$) as that of HD~26965 (Fe/H$=-0.42\pm0.04$), which 
is consistent with the finding of \citet{petigura18} that smaller planets are detected around stars with
wide-ranging metallicities. 

Based on the observed properties of HD~26965~b, several inferences of 
the planet's properties and history are possible. With a minimum mass of 
8.4$M_{\earth}$, the planet likely possesses a gaseous atmosphere based on 
other planets with known masses and radii \citep{rogers15}. However, we 
note that Kepler-10 c has a similar mass and orbit, is hosted by a similar, 
low-metallicity star \citep{bat11, fre11, weiss16, rajpaul17}, and does not possess an 
envelope \citep{lopez14}, so HD~26965~b may be a similar type of world. 
In the near-term, this possibility can only be resolved if a transit is 
detected.

\begin{figure}
\includegraphics[angle=0,width=8.2cm]{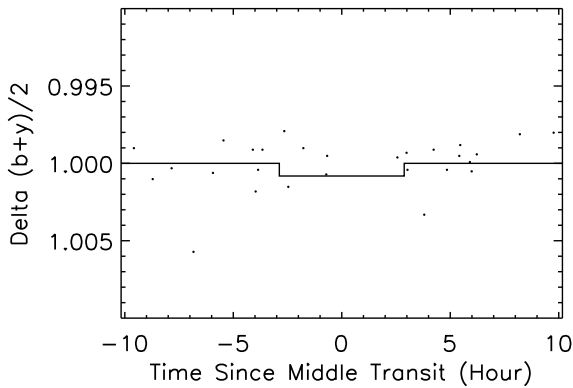}
\caption{ Phased photometric data of HD~26965 from Fairborn Observatory. It is an expanded version of Figure.~\ref{fig:photo}. 
This figure is centered on the predicted middle transit time. The solid curve shows the predicted central transit signal with a depth of 0.0008~mag. 
The current data do not support the existence of a transit signal.  \label{fig:photo2}
} 
\end{figure}

Detecting exoplanets via the RV technique and subsequently monitoring their transit windows is one of the most fruitful strategies for finding bright stars with transiting planets, which are the best candidates for exoplanet atmospheric studies. Figure~\ref{fig:photo2} shows the photometric data near the predicted middle transit time. The expected duration of a central transit is $\sim$$R_{\star}P/(\pi a)=5.8$~hr, and the expected depth is $(R_{b}/R_{\star})^2$$\sim$0.0008. Our current photometric data do not support the existence of a shallow transit from HD~26965b. 


Lastly, the detection of HD~26965b shows the advantage of the Dharma planet survey strategy. 
The high precision and high cadence RV campaign from TOU have 
greatly increased the detection sensitivity of low-mass planets. The fact 
that we can discover this system with similar RV precision to HAPRS (Ma \& Ge, in preparation), but with high cadence 
(133 nights observations within 2 years using TOU versus 97 nights observations within 13 years using HARPS), 
demonstrates that high precision and high cadence RV surveys of bright stars 
in the solar neighborhood will likely lead to the detection of a large number of low-mass planets with high completeness, and 
possible detections of low-mass planets in their habitable zones \citep{ge16}.

\section*{Acknowledgements}
We thank the anonymous referee for comments and suggestions which have helped significantly improve the quality of this paper. We are grateful to the Dharma Endowment Foundation for generous support.  We thank Indiana University for 
the donation of their 50-inch telescope to the Department of Astronomy, University of Florida. B.M thanks the support of the 
NASA-WIYN observation award. J.I.G.H. acknowledges financial support from the Spanish Ministry of Economy and 
Competitiveness (MINECO) under the 2013 Ram\'on y Cajal program MINECO RYC-2013-14875, and the Spanish ministry project MINECO AYA2014-56359-P.
We are grateful to Lou Boyd of Fairborn Observatory, and engineering staff at 
Steward Observatory, Bruce Hille, Scott Swindell, Joe Horscheidt, Melanie Waidanz, Chris Johnson,  
and Jeff Kingsley for providing excellent engineering support for the Dharma Planet Survey. 
This paper is dedicated to the memory of the wonderful technical manager of Steward Observatory, 
Mr. Robert Peterson, who helped get the DEFT project going on Mt. Lemmon in 2015-2016 before he passed away on October 20, 2016. 
Greg Henry acknowledges long-term support from NASA, NSF, Tennessee State 
University, and the State of Tennessee through its Centers of Excellence program.







%
%
%


\bsp	
\label{lastpage}
\end{document}